\documentclass{article}
\usepackage{amsmath}
\usepackage{amssymb}
\usepackage{epsfig,amsmath,amssymb,bm,epsf,graphics}

\def\lsim{\lower.5ex\hbox{$\; \buildrel < \over \sim \;$}}
\def\gsim{\lower.5ex\hbox{$\; \buildrel > \over \sim \;$}}
\def\AHR{analogue Hawking radiation}
\def\AH{acoustic horizon}

\input{tcilatex}

\begin{document}

\title{Analogue Hawking Radiation from Astrophysical Black Hole Accretion}
\author{Tapas K. Das \\ \\
N. Copernicus Astronomical Center\\
Bartycka 18 00-716 Warsaw, Poland\\
{\it tapas@camk.edu.pl}\\
}
\maketitle

\begin{abstract}
We show that spherical accretion onto astrophysical black holes
can be considered as a natural example of analogue system.
We provide, for the first time, an exact analytical
scheme for calculating the analogue Hawking temperature and surface
gravity for general relativistic accretion onto astrophysical black
holes. Our calculation may bridge the gap between the theory of
transonic astrophysical accretion and the theory of analogue Hawking
radiation. We show that the domination of the analogue Hawking temperature
over the actual Hawking temperature may be a real astrophysical phenomena.
We also discuss the possibilities of the emergence of analogue white holes
around astrophysical black holes. Our calculation is general enough to
accommodate accreting black holes with any mass.
\end{abstract}

\section{Introduction}
\noindent
By providing an analogy between the propagation of light in 
a curved geometry and the acoustic propagation in a 
steady, inviscid, transonic barotropic fluid, it has been 
demonstrated \cite{unruh1,jacob1,unruh2,visser,jacob2,bilic} 
that the representative fluid posses the
signature of an acoustic black hole (BH hereafter) embedded by an acoustic horizon
at its transonic point (where the fluid Mach number 
becomes unity).
This acoustic horizon
emits thermal radiation known as analogue Hawking radiation.
The temperature of such radiation is
called analogue Hawking temperature, or acoustic Hawking temperature
(Unruh \cite{unruh1} was the first to propose and compute such a
temperature). 
Throughout our work, we use $T_{{AH}}$
to denote the analogue Hawking temperature, and $T_H$ to denote the
the actual Hawking temperature
where $T_H={\hbar{c^3}}/{8{\pi}{k_b}GM_{BH}}$,
$M_{BH}$ being the mass of the black hole. 
We use the words `analogue', `acoustic' and `sonic' synonymously
in describing the horizons or black holes.
The theory of {\AHR}  opens up the possibility to experimentally verify some 
basic features of BH physics by creating the sonic horizons in the 
laboratory. A number of works have been carried out to formulate the
condensed matter or optical analogs of event horizons
\cite{arti}. It is also 
expected that {\AHR}  may find important uses in the fields of inflationary 
models, quantum gravity and sub-Planckian models of string theory
\cite{peran}. \\
\noindent
Since the publication of the seminal paper by Bondi in 1952
\cite{bondi}, the transonic 
behaviour of accreting fluid onto compact astrophysical objects has 
been extensively studied in the astrophysics community (including the 
fast ever paper on curved acoustic geometry by Moncrief \cite{moncr}), and the 
pioneering work by Unruh in 1981 \cite{unruh1}
initiated a substantial number of works
in the theory of {\AHR}  with diverse fields of application stated above. 
No attempt has yet been made to bridge
these two categories of research, astrophysical BH accretion
and the theory of {\AHR},  by providing a self-consistent study of {\AHR}  for real 
astrophysical fluid flows, i.e., by establishing the fact that
accreting black holes can be considered as a natural 
example of analogue system. Since both the theory of transonic
astrophysical accretion and the theory of 
{\AHR}  stem from almost 
exactly the same physics, the propagation of a transonic fluid with 
acoustic disturbances embedded into it, it is important to study 
{\AHR} for transonic accretion onto astrophysical black 
holes and to compute $T_{{AH}}$ for such accretion. \\
In this paper we show that a spherically accreting black hole 
naturally provides a concrete example of analogue system 
found in nature.
We consider general 
relativistic (GR) spherically symmetric accretion of inviscid 
polytropic fluid onto a Schwarzschild BH of mass $M_{BH}$. We first formulate 
and solve the conservation equations for such accretion and then 
demonstrate that such flows may become transonic depending on initial boundary 
conditions. We then, for the first time
\footnote{Note that Unruh's original calculation of $T_{{AH}}$ \cite{unruh1} assumes the 
positional constancy of sound speed, which may not always 
be a real physical situation.
This was modified to include the position dependent sound speed by 
Visser \cite{visser} for flat space and by Bili$\acute{c}$ \cite{bilic} 
for curved space. However, none of these 
papers provides the exact calculation of $r_h$ and {\AH} related quantities
from  first principles, and hence the numerical value of $T_{{AH}}$ in {\it any}
existing literature (even for flat space, let alone curved space) is 
obtained using approximate, order of magnitude, calculations only.}, 
{\it completely  analytically} calculate the {\it exact} location of
the acoustic horizon $r_h$ \cite{tapas} and calculate the 
quantities on the {\AH} which are required to compute the surface gravity and $T_{{AH}}$
for our model.
We then analytically 
calculate the exact value of $T_{{AH}}$ in {\it curved space time} for
{\it all possible}
real physical spherical accretion solutions. \\
It is important to
note that the accreting astrophysical BHs are the {\it only} 
real physical candidates for which both the horizons, actual (electromagnetic)
and analogue (sonic), may exist together {\it simultaneously}. Hence 
our application of {\AHR} to the theory of transonic 
astrophysical accretion may be useful to compare 
some properties of these two kind of horizons,
by theoretically studying and comparing the behaviour of the {\it same} flow
close to these horizons. As
an illustrative example of such a procedure, we also calculate 
and compare $T_{{AH}}$ and $T_H$ for the same BH and for the same set of initial
boundary conditions, and show that for astrophysically interesting regions of
parameter space, $T_{{AH}}$ may well exceed $T_H$.
One should, however, note that both $T_H$ and $T_{AH}$ comes out to be
quite less compared to the macroscopic classical fluid temperature
of accreting matter. Hence BH accretion system may not be a good
candidate to allow any observational test for detecting the analogue 
radiation. Our work rather introduces the most relevant classical system 
to manifest such analogues in curved space-time.
Also we
find the possibility that an actual BH event horizon may be embedded by an
acoustic white hole. 
\\
\section{Formalism and results}
We take the Schwarzschild radius $r_g=2GM_{BH}/c^2$, and scale the radial 
coordinate $r$ in units of $r_g$, any velocity in units of $c$, and all other
physical quantities are scaled accordingly. We set $G=c=M_{BH}=1$. The mass of the 
accreting fluid is assumed to be much less compared to $M_{BH}$ (which is usually
reality for astrophysical BH accretion), so that the gravity field is 
controlled essentially by $M_{BH}$ only. Accretion is 
$\theta$ and $\phi$ symmetric and posses only radial inflow velocity. We 
concentrate on stationary solutions. Accretion is governed by the radial part
of the GR time independent Euler and continuity equations in Schwarzschild
metric. The conserved specific flow energy ${\cal E}$ (the relativistic 
analogue of Bernoulli's constant) along each stream line reads ${\cal E}=hu_t$ 
\cite{anderson} where
$h$ and $u_\mu$ are the specific enthalpy and the four velocity, which can be 
re-cast in terms of the radial three velocity $u$ and the polytropic sound speed $a$ to obtain:
$$
{\cal E}=\left[\frac{\gamma-1}{\gamma-\left(1+a^2\right)}\right]
\sqrt{\frac{1-\frac{1}{r}}{1-u^2}}.
\eqno{(1)}
$$
In this work we concentrate on positive Bernoulli constant solutions.
The mass accretion rate ${\dot M}$ may be obtained by integrating the continuity
equation:
$$
{\dot M}=4{\pi}{\rho}ur^2\sqrt{\frac{r-1}{r\left(1-u^2\right)}},
\eqno{(2)}
$$
where $\rho$ is the proper mass density.  A polytropic equation of state, $p=K\rho^\gamma$,  is 
assumed (this is common in astrophysics
to describe relativistic accretion) to define another constant $\Xi$ as:
$$
\Xi=K^{\frac{1}{1-\gamma}}{\dot M}=4{\pi}{\rho}ur^2\sqrt{\frac{r-1}{r\left(1-u^2\right)}}
\left[\frac{a^2\left(\gamma-1\right)}{\gamma-\left(1+a^2\right)}\right].
\eqno{(3)}
$$
Simultaneous solution of eq.\ (1--3) provides the dynamical three velocity gradient 
at any radial distance $r$:
$$
\frac{du}{dr}=\frac{u\left(1-u^2\right)\left[a^2\left(4r-3\right)-1\right]}
{2r\left(r-1\right)\left(u^2-a^2\right)}=\frac{{\cal N}}{{\cal D}}.
\eqno{(4)}
$$
The denominator 
of eq.\ (4) becomes zero at $r=1$ (actual horizon) and when $u=a$ (acoustic
horizon). To maintain the physical smoothness of the flow for any $r>1$, the numerator
${\cal N}$ and the denominator ${\cal D}$
of eq.\ (4) have to simultaneously vanish. Hence by making ${\cal N}$ = ${\cal D}$ = 0,
we get the so-called sonic point conditions (value of dynamical and sound velocities
in terms of horizon location $r_h$) as
$$
u_h=a_h=\sqrt{\frac{1}{4r_h-3}}.
\eqno{(5)}
$$
Substitution of $u_h$ and $a_h$ into eq.\ (1) for $r=r_h$ provides:
$$
r_h^3+r_h^2\Gamma_1+r_h\Gamma_2+\Gamma_3=0,
\eqno{(6)}
$$
where
$$
\Gamma_1=\left[\frac{2{\cal E}^2\left(2-3\gamma\right)+9\left(\gamma-1\right)}
         {4\left(\gamma-1\right)\left({\cal E}^2-1\right)}\right],~
$$
$$
\Gamma_2=\left[\frac{{\cal E}^2\left(3\gamma-2\right)^2-
          27\left(\gamma-1\right)^2}
          {32\left({\cal E}^2-1\right)\left(\gamma-1\right)^2}\right],~
\Gamma_3=\frac{27}{64\left({\cal E}^2-1\right)}.
\eqno{(7)}
$$
Solution of eq.\ (6) provides the location of the {\AH} in terms of only two accretion parameters
$\{{\cal E},\gamma\}$, which is our two parameter input set to study the flow,
and hereafter
will be denoted by ${\cal P}_2$. We solve eq. (6) completely analytically 
by employing the Cardano-Tartaglia-del Ferro technique. We define:
$$
\Sigma_1=\frac{3\Gamma_2-\Gamma_1^2}{9},~
\Sigma_2=\frac{9\Gamma_1\Gamma_2-27\Gamma_3-2\Gamma_1^3}{54},~
\Psi=\Sigma_1^3+\Sigma_2^2,~
$$
$$
 \Theta={\rm cos}^{-1}\left(\frac{\Sigma_2}{\sqrt{-\Sigma_1^3}}\right)
\Omega_1=\sqrt[3]{\Sigma_2+\sqrt{\Sigma_2^2+\Sigma_1^3}},~
$$
$$
\Omega_2=\sqrt[3]{\Sigma_2-\sqrt{\Sigma_2^2+\Sigma_1^3}},~
\Omega_{\pm}=\left(\Omega_1\pm\Omega_2\right)
\eqno{(8)}
$$
so that the three roots for $r_h$ come out to be:
$$
^1\!r_h=-\frac{\Gamma_1}{3}+\Omega_+,~
^2\!r_h=-\frac{\Gamma_1}{3}-\frac{1}{2}\left(\Omega_+-i\sqrt{3}\Omega_-\right),~
$$
$$
^3\!r_h=-\frac{\Gamma_1}{3}-\frac{1}{2}\left(\Omega_--i\sqrt{3}\Omega_-\right).
~\eqno{(9)}
$$
However, note that not all $^i\!r_h\{i=1,2,3\}$ would be real for all ${\cal P}_2$. It is
easy to show that if $\Psi>0$, only one root is real; if $\Psi=0$, all roots are
real and at least two of them are identical; and if $\Psi<0$, all roots are real 
and distinct. Selection of the real physical ($r_h$ has to be greater than unity) roots
requires
the following discussion. \\ 
Although we analytically calculate $r_h$ and other variables at $r_h$, there is no 
way that one can analytically calculate the flow variables at any arbitrary $r$. One needs
to integrate eq.\ (6) numerically to obtain the transonic profile of the flow for 
all range of $r$, starting from infinity and ending on to the actual event horizon. To do
so, one must set the appropriate limits on ${\cal P}_2$ to model the realistic situations
encountered in astrophysics. As ${\cal E}$ is scaled in terms of
the rest mass energy and includes the rest mass energy (${\cal E}=1$ corresponds to
rest mass energy), ${\cal E}<1$ corresponds to the negative energy accretion state where
radiative extraction of rest mass energy from the fluid is required. For such extraction 
to be made possible, the accreting fluid has to
posses viscosity or other dissipative mechanisms, which may violate the Lorenzian invariance.
On the other hand, although almost any ${\cal E}>1$ is mathematically allowed, large 
values of ${\cal E}$ represents flows starting from infinity 
with extremely high thermal energy, and ${\cal E}>2$ accretion represents enormously 
hot (thermal energy greater than the rest mass energy) initial ($r\rightarrow{\infty}$)
flow configurations, which are not properly conceivable in realistic astrophysical situations.
Hence we set $1{\lsim}{\cal E}{\lsim}2$. Now, $\gamma=1$ corresponds to isothermal accretion
where accreting fluid remains optically thin. This is the physical lower limit for 
$\gamma$; $\gamma<1$ is not realistic in accretion
astrophysics. On the other hand,
$\gamma>2$ is possible only for superdense matter 
with substantially large magnetic
field (which requires the accreting material to be governed by GR magneto-hydrodynamic 
equations, dealing with which
is beyond the scope of this paper) and direction dependent anisotropic pressure. We thus 
set $1{\lsim}\gamma{\lsim}2$ as well, so ${\cal P}_2$ has the boundaries
$1{\lsim}\{{\cal E},\gamma\}{\lsim}2$. However, one should note that the most preferred 
values of $\gamma$ for realistic BH accretion ranges from $4/3$ 
to $5/3$ \cite{frank}.\\
\noindent
Coming back to our solution for $r_h$, we find that for our preferred range of ${\cal P}_2$,
one {\it always} obtains $\Psi<0$. Hence the roots are always real and three real
unequal roots can be computed as:
$$
^1\!{{r}}_h=2\sqrt{-\Sigma_1}{\rm cos}\left(\frac{\Theta}{3}\right)
                  -\frac{\Gamma_1}{3},
^2\!{{r}}_h=2\sqrt{-\Sigma_1}{\rm cos}\left(\frac{\Theta+2\pi}{3}\right)
                  -\frac{\Gamma_1}{3},~
$$
$$
^3\!{{r}}_h=2\sqrt{-\Sigma_1}{\rm cos}\left(\frac{\Theta+4\pi}{3}\right)
                  -\frac{\Gamma_1}{3}
\eqno{(10)}
$$
We find that for {\it all} $1{\lsim}{\cal P}_2{\lsim}2$, $^2\!{{r}}_h$ becomes negative. We 
then find that $\{^1\!{{r}}_h,^3\!{{r}}_h\}>1$ for most values of our astrophysically 
tuned ${\cal P}_2$.
However, it is also found that $^3\!{{r}}_h$ does not allow steady physical flows to pass
through it; either $u$, or $a$, or both, becomes superluminal before the flow reaches
the actual event horizon, or the Mach number profile shows intrinsic fluctuations for 
$r<r_h$. We obtain this information by numerically integrating the 
complete flow profile passing through $^3\!{{r}}_h$. Hence it turns out that we need to
concentrate {\it only} on  $^1\!{{r}}_h$ for realistic astrophysical BH accretion. 
Both large ${\cal E}$ and large $\gamma$ enhance the thermal energy of the flow 
so that the 
accreting fluid acquires the large radial velocity to overcome $a$ only in the 
close vicinity of the BH. Hence $r_h$ anti-correlates with ${\cal P}_2$. 
To obtain
$(dv/dr)$ and $(da/dr)$ on the {\AH}, we apply L' Hospital's rule to 
eq.\ (4) and we have:
$$
\left(\frac{du}{dr}\right)_{r=r_h}=\Phi_{12}-\Phi_{123},~
$$
$$
\left(\frac{da}{dr}\right)_{r=r_h}=\Phi_4\left(\frac{1}{\sqrt{4r_h-3}}+\frac{\Phi_{12}}{2}
                            -\frac{\Phi_{123}}{2}\right)
\eqno{(11)}
$$
where
$$
\Phi_{12}=-\Phi_2/2\Phi_1,~
\Phi_{123}=\sqrt{\Phi_2^2-4\Phi_1\Phi_3}/2\Phi_1,~
$$
$$
\Phi_1=\frac{6r_h\left(r_h-1\right)}{\sqrt{4r_h-3}},~
\Phi_2=\frac{2}{4r_h-3}\left[4r_h\left(\gamma-1\right)-
        \left(3\gamma-2\right)\right]
$$
$$
\Phi_3=\frac{8\left(r_h-1\right)}{\left(4r_h-3\right)^{\frac{5}{2}}}
       [r_h^2\left(\gamma-1\right)^2-r_h\left(10\gamma^2-19\gamma+9\right)
$$
$$
       +\left(6\gamma^2-11\gamma+3\right)],~
\Phi_4=\frac{2\left(2r_h-1\right)-\gamma\left(4r_h-3\right)}
       {4\left(r_h-1\right)}
\eqno{(12)}
$$
The GR description of the sonic geometry for a
generalized acoustic metric is provided by Bili$\acute{c}$ \cite{bilic}. 
Keeping in the mind that the 
surface gravity can be obtained by computing the norm of the gradient of the norm of the 
Killing fields evaluated at $r_h$, we use the methodology developed in 
Bili$\acute{c}$ \cite{bilic} to calculate 
the general relativistic expression for surface gravity, and $T_{{AH}}$ for our flow geometry
is obtained as:
$$
T_{{AH}}=\frac{\hbar{c^3}}{4{\pi}{k_b}GM_{BH}}
    \left[\frac{r_h^{\frac{1}{2}}\left(r_h-0.75\right)}
    {\left(r_h-1\right)^{\frac{3}{2}}}\right]
    \left|\frac{d}{dr}\left(a-u\right)\right|_{r=r_h},
\eqno{(13)}
$$
where the values of $r_h, (du/dr)_h$ and $(da/dr)_h$ are obtained using the system of
units and scaling used in our paper.  The stationarity approximation allows us not to discriminate
between the acoustic apparent and event horizon. Note again, that, since $r_h$ and other quantities
appearing in eq.\ (13) are {\it analytically} calculated as a function of
${\cal P}_2$, we provide an {\it exact analytical value} of the 
general relativistic analogue Hawking temperature for {\it all possible solutions} of an 
spherically accreting
astrophysical black hole system, something which has never been done in the literature.
If $\sqrt{4r_h-3}(1/2-1/\Phi_4)(\Phi_{12}-\Phi_{123})>1$,
one {\it always} obtains $(da/dr<du/dr)_h$, which indicates the presence of the
acoustic white holes at $r_h$. This inequality holds good for
certain astrophysically relevant range of ${\cal P}_2$; see following discussions.\\
For a particular value of ${\cal P}_2$, we now define the quantity $\tau$ to be the ratio of
$T_{{AH}}$ and $T_H$, i.e., $\tau=T_{{AH}}/T_H$.
This ratio $\tau$ comes out to be independent of the mass of the black hole, 
which enables us to compare the properties of two kind of horizons (actual and acoustic) for
an spherically 
accreting BH with any mass, 
starting from the primordial holes
to the super massive black holes
at galactic centers. Note, however, that we applied the analogue to 
describe the classical perturbation of the fluid in terms of a 
field satisfying the wave equation on a curved effective geometry. Our
main motivation in this paper is not to rigorously demonstrate how
the phonon field generated in this system could be made quantized. To
accomplish that task, one can show that the effective action for the
acoustic perturbation is equivalent to the field theoretical action
in curved space, and the corresponding commutation and dispersion 
relations (see, e.g., \cite{unruh3}) may directly follow from there. However, such calculations
are beyond the scope of this paper and would be presented elsewhere.\\
\begin{figure}
\vbox{
\vskip -2.5cm
\centerline{
\psfig{file=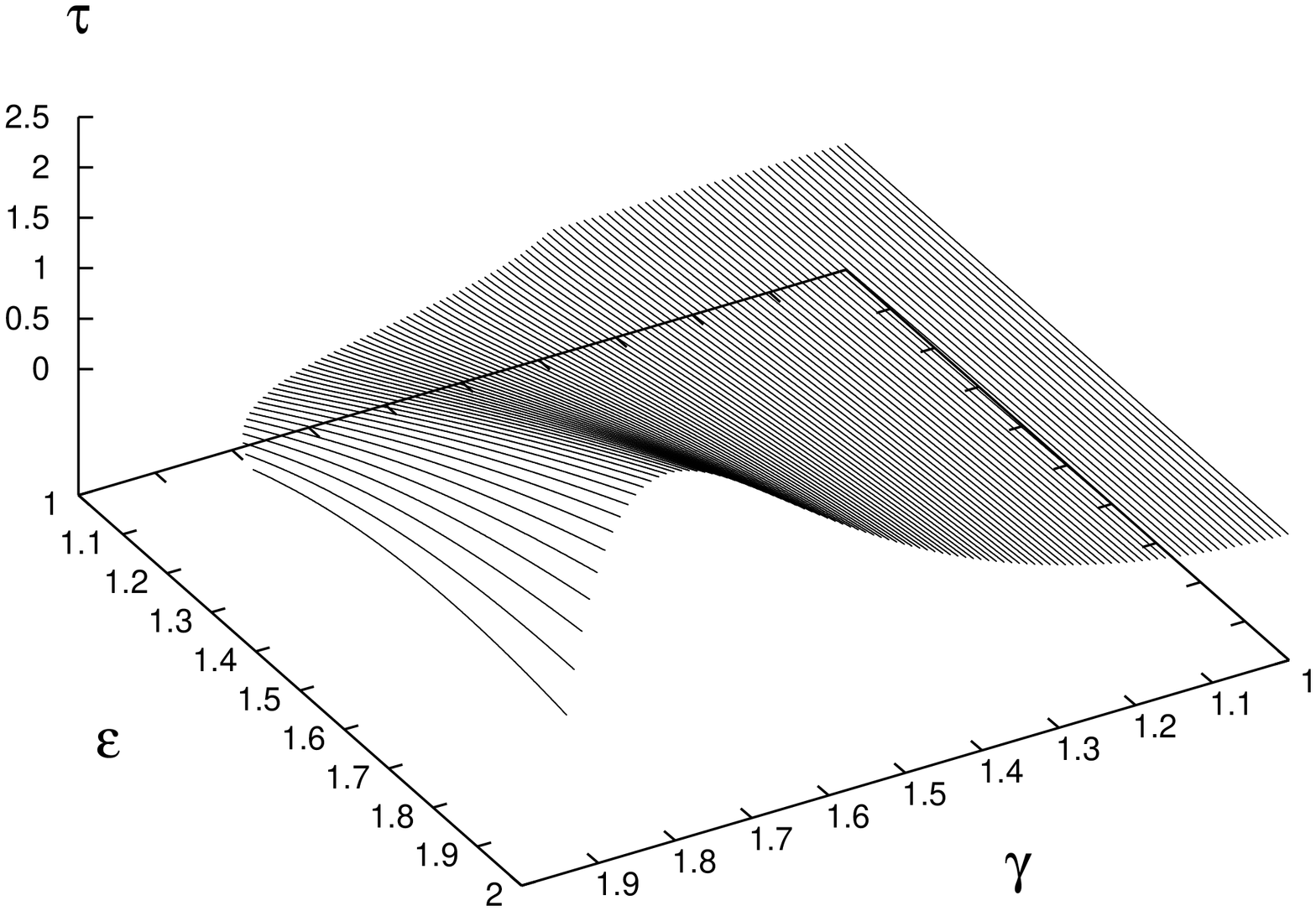,height=12cm,width=10cm,angle=270.0}}}
\noindent {{\bf Fig. 1:}
Variation of ratio of the analogue to the actual
Hawking temperature ($\tau$) on conserved specific energy (${\cal E}$)
and polytropic indices ($\gamma$) of the flow.}
\end{figure}
\begin{figure}
\vbox{
\vskip -4.5cm
\centerline{
\psfig{file=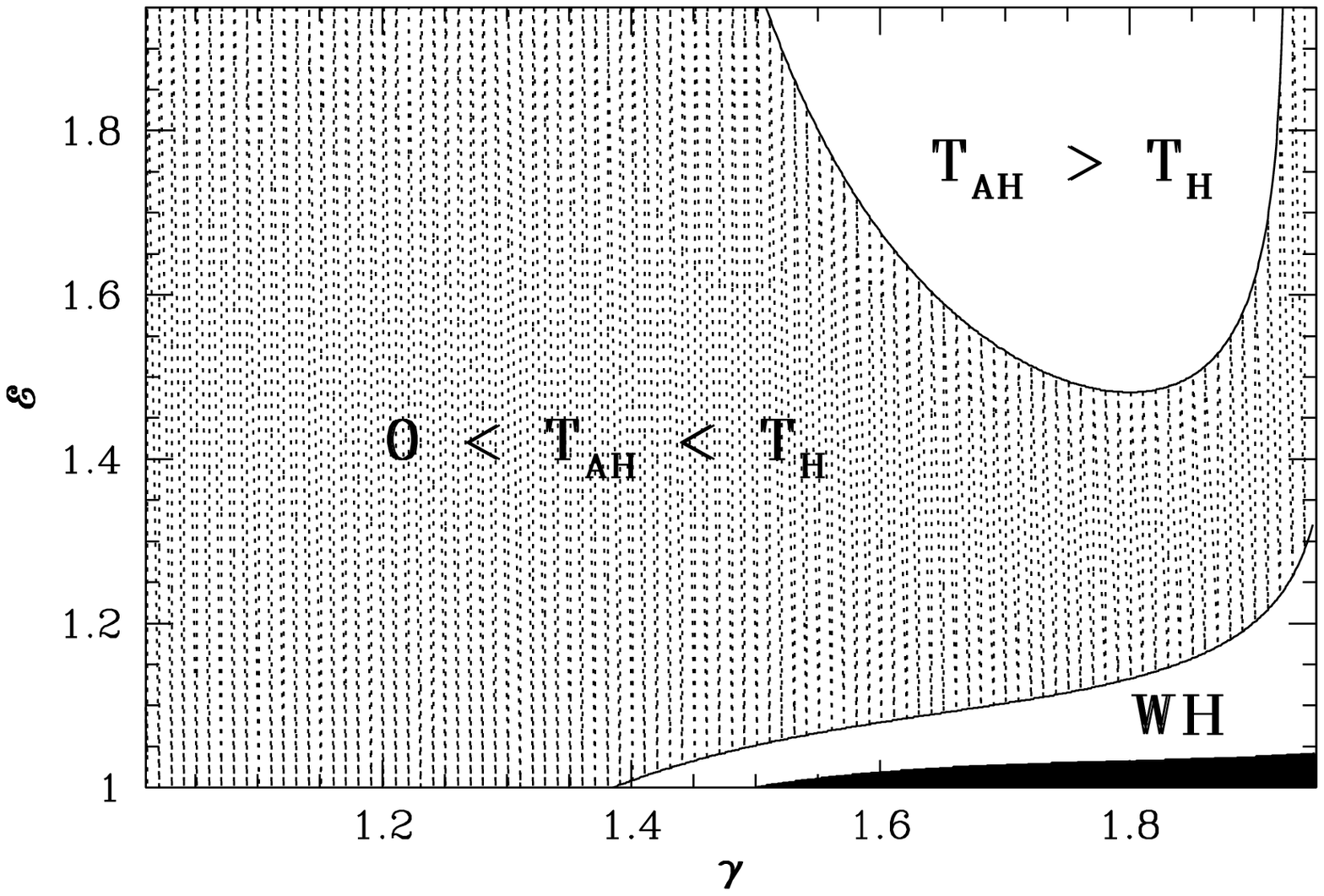,height=11cm,width=11cm,angle=0.0}}}
\noindent {{\bf Fig. 2:}
Parameter space classification for ${\cal P}_2$ for four different sets of values
of $T_{{AH}}$.}
\end{figure}
\noindent
Figure 1 shows the dependence of $\tau$ on ${\cal P}_2$. The dependence is highly nonlinear.
We plot $\tau$ corresponding to the black hole solutions only (solutions for which
$(da/dr)_h>(du/dr)_h$).
Note that $T_{{AH}}>T_H$ for
some ${\cal P}_2$. In figure 2, we scan the complete ${\cal P}_2$ space by depicting its
four important non-overlapping regions.
For accretion with ${\cal P}_2$ taken from the lightly shaded 
regions marked by $0<T_{{AH}}<T_H$, 
the Hawking temperature dominates over its analogue counterpart.
For high $\gamma$, high ${\cal E}$ flow (`hot' accretion),
{\AHR} becomes the dominant process compared to the actual Hawking radiation 
($T_{{AH}}>T_H$).
For low ${\cal E}$ and intermediate/high $\gamma$, acoustic white holes appear
(white region marked by {\bf WH}, where ${da/dr<du/dr}$ at $r_h$). 
For ${\cal P}_2$ belonging to this region, one obtains outflow (outgoing solutions with Mach number 
increasing with increase of $r$) only. This has also been verified by obtaining the 
complete flow profile by integrating eq.\ (6).
The dark shaded 
region (lower right corner of the figure) represents ${\cal P}_2$ for which $r_h$ comes out
to be physical ($r_h>1$) but
$\Phi_{123}^2$ becomes negative, hence $(du/dr)_h$ and $(da/dr)_h$ are not real and $T_{{AH}}$
becomes imaginary. Note that both $T_{{AH}}>T_H$ and white hole regions are obtained even for 
$4/3<\gamma<5/3$, which is the range of values of $\gamma$ for most realistic flows of matter 
around astrophysical black holes.
{\it Hence we have shown that the domination of
the analogue Hawking temperature over the actual Hawking temperature
and the emergence of 
analogue white holes are real astrophysical phenomena}. 
We did not perform formal stability
analysis on white hole solutions. It may be 
possible that the white hole solutions are 
unstable,
according to the recent proposal that the sonic white holes posses intrinsic instability \cite{leon}.
However, such detailed investigations
are beyond the scope of this paper and we would like to defer it for our future work.\\
\section{Discussion}
In this paper we have studied spherically symmetric accretion onto astrophysical 
black holes as a natural example of analogue system. If accreting matter contains non-zero
intrinsic angular momentum density, accretion becomes axisymmetric and forms 
disc like structure surrounding the black hole.
Calculations presented in this paper 
may be extended for astrophysical BH accretion
discs as well. Bili$\acute{c}$ \cite{bilic} studied the relativistic acoustic geometry for 
2D axisymmetric space time
and one can use such formalism to study {\AHR} for GR and post-Newtonian multi-transonic, rotating,
advective BH accretion discs in the similar way as is done in this letter. 
Viscosity, however, is quite a subtle issue in studying the analogue effects for 
disc accretion. 
From the astrophysical point of view, even thirty years after the discovery of 
standard accretion disc theory, exact modeling of viscous transonic BH accretion, including 
proper heating and cooling mechanisms is still quite an arduous task, even for Newtonian flow,
let alone for general relativistic accretion. And from analogue model point of view, viscosity
is likely to destroy the Lorenzian invariance, and hence the assumptions behind building up an
analogue model may not remain that much consistent. 
However, there are two possibilities through which the calculations presented in the paper 
may be translated onto the idea of disc accretion. Firstly,
one can study the disc system in its weak viscosity limit.
Secondly, if viscosity is present,
one can calculate the ratio of the in-fall to viscous time scale to see at which length scale
the ratio comes out to be less than one. At that length scale, accretion acquires such a high
radial in-fall velocity (measured on the equatorial plane of the disc) that the residence 
time of matter at a particular annulus of the disc
becomes quite low and hence the angular 
momentum transport may become inefficient. So the flow may practically be taken as inviscid
in that length scale. Now if it is found that the length scale at which the sonic horizon
is formed, is more or less comparable or less than the length scale where viscosity becomes
insignificant, then it should not be any problem in studying the analogue model even for disc
accretion. The whole parameter space for which one can model the analogue process using 
accretion disc, will then depend on the accretion parameters for which 
the acoustic horizon forms at a distance equal to or less than the radial distance
where the in-fall time scale is smaller compared to the viscous time scale.
Our preliminary calculations show that {\it indeed} there are situations for which the 
above condition is self-consistently satisfied.
Such calculations
are in progress and will be presented elsewhere.

\textit{Acknowledgments}
I am deeply indebted to Neven Bili$\acute{c}$ for his help during my process
of understanding the analogue Hawking radiation in curved geometry. 
I am grateful to William Unruh, Ted Jacobson, John Miller and
Stefano Liberati 
for their insightful suggestions provided after going through
the first draft of this manuscript, which have been useful to modify the 
manuscript to this present form. I 
acknowledge stimulating discussions with 
Paul Wiita and Ralf Sch$\ddot{\rm u}$tzhold. 
This work is supported by grants from the KBN.

\end{document}